\documentclass[pra,twoside,showpacs,superscriptaddress,twocolumn]{revtex4}

\usepackage{psfig,latexsym}

\begin{document}
\title{Several experimental realizations of symmetric phase-covariant quantum cloner of
single-photon qubits}

\author{Jan Soubusta}
\affiliation{Joint Laboratory of Optics of Palack\'{y} University and
     Institute of Physics of Academy of Sciences of the Czech Republic,
     17. listopadu 50A, 779\,07 Olomouc, Czech Republic}

\author{Lucie Bart\r{u}\v{s}kov\'{a}}
\affiliation{Department of Optics, Palack\'y University,
     17.~listopadu 50, 772\,00 Olomouc, Czech~Republic}

\author{Anton\'{\i}n \v{C}ernoch}
\affiliation{Joint Laboratory of Optics of Palack\'{y} University and
     Institute of Physics of Academy of Sciences of the Czech Republic,
     17. listopadu 50A, 779\,07 Olomouc, Czech Republic}

\author{Jarom{\'\i}r Fiur\'{a}\v{s}ek}
\affiliation{Department of Optics, Palack\'y University,
     17.~listopadu 50, 772\,00 Olomouc, Czech~Republic}

\author{Miloslav Du\v{s}ek}
\affiliation{Department of Optics, Palack\'y University,
     17.~listopadu 50, 772\,00 Olomouc, Czech~Republic}

\date{\today}

\begin{abstract}
We compare several optical implementations of phase-covariant cloning machines. 
The experiments are based on copying of the polarization state of a single photon 
in bulk optics by special unbalanced beam splitter or by balanced beam splitter 
accompanied by a state filtering. Also the all-fiber based setup is discussed, 
where the information is encoded into spatial modes, i.e., the photon can
propagate through two optical fibers. Each of the four implementations possesses
some advantages and disadvantages that are discussed.

\end{abstract}


\maketitle

%%%%%%%%%%%%%%%%%%%%%%%%%%%%%%%%%%%%%%%%%%%%%%%%%%%%%%%%%%%%%%%%%%%%%%%%%%%%%%%%%%%%%%%%%%%%%%%%%%%%

\section{Introduction}

One of the most fundamental consequences of the laws of
quantum mechanics is the impossibility of an exact copying
of an unknown quantum state \cite{Wootters82}.
Nevertheless, this no-go statement does not exclude a
possibility of an approximate copying \cite{Buzek96}. 
In general, quantum cloning machines
can create $M$ approximate clones from $N<M$ originals. The
simplest ones produce two approximate copies from one original
unknown qubit state using one ancilla qubit. The
quality of cloning can be well quantified by the measure of
fidelity, which is defined as the overlap of each copy with
the original input state. Optimal cloners maximize
average fidelities of the clones. In the symmetric case the
fidelities of all clones are equal. Asymmetric cloning
transformations allow different fidelities of particular
clones, i.e., smaller distortion of some copies at the cost
of lower precision of the others.  The theory of optimal quantum cloners 
is well established and optimal cloning transformations 
are known for many classes of input states \cite{Scarani05,Cerf06}.

In optical realizations of quantum information processing, one 
is mostly interested in cloning of the states of single photons. 
Most experimental realizations up to now implemented a universal cloning
transformation using either stimulated parametric down-conversion
\cite{Linares02,Fasel02,DeMartini04} or interference 
of photons on balanced beam splitter \cite{Ricci04,Irvine04,Khan04}. 
In the case of universal cloner, fidelities of the clones do not depend 
on the input state to be copied \cite{Buzek96,Gisin97}. Sometimes, however,
we want to clone only a certain subset of states and in this case a higher
fidelity may be obtained using an appropriately  tailored cloner. In particular, 
the \emph{phase-covariant} cloning machine optimally clones all qubit states 
from the equator of the Bloch sphere \cite{Fuchs97,Soub04,Du05,Sciarrino05}. 
More precisely, fidelities of cloned qubit 
states produced by a phase-covariant cloning transformation do not depend on the
mutual phase between amplitudes of two fixed-basis states 
$|0\rangle$ and $|1\rangle$ but depend on
their ``intensity'' ratio. For a subset of states with a
fixed ``intensity'' ratio the optimal phase-covariant cloner
offers higher fidelities of clones than the universal one.
Phase-covariant cloner is of great interest also because it
can be used for an optimal individual attack on BB84 quantum
key distribution protocol \cite{Scarani05}.

In this paper we describe several different experimental
realizations of the optimal symmetric phase-covariant
$1\to2$ cloning of optical qubit states and compare their performances.
The  schemes that we consider can be broadly divided into two
categories. The first approach to cloning of polarization states of 
single photons relies on the two-photon interference on a specially 
tailored unbalanced beam splitter which exhibits different transmittance for
vertical and horizontal polarizations \cite{Fiurasek03}. Besides the scheme 
utilizing a special custom-made beam splitter that was described in our 
earlier publication \cite{Cernoch06}, we also present and investigate 
a setup where the unbalanced beam splitter is emulated by a Mach-Zehnder 
interferometer \cite{Zhao05} with Soleil-Babinet compensator in each arm.

We also propose and demonstrate a second alternative approach to optimal 
phase-covariant cloning which combines the bunching of photons on 
a balanced beam splitter followed by a state-dependent filtering operation. 
With this latter method we were able to demonstrate high-quality phase-covariant 
cloning of polarization states of single photons, with average fidelities 
exceeding the limit of optimal universal cloning. 

For completeness, we also briefly review the all-fiber based cloning scheme, 
where the qubits are encoded in a state of a single photon which can propagate 
in two different single-mode optical fibers \cite{Bartuskova06}. This scheme 
represents again the first cloning approach, where two variable ratio couplers 
stand in for the special beam splitter. With this last setup we were able 
to accomplish high-quality phase-covariant cloning of single-photon states 
encoded into spatial modes. The average fidelities exceed the limit of 
optimal universal cloning. 

The article is organized as
follows.  Section \ref{Sec_Theory} recapitulates briefly the
basic theoretical tools and describes theoretically the
different possible implementations of a phase-covariant
cloning machine. Section \ref{sec_MZ} deals with an
implementation employing a Mach-Zehnder interferometer in
the role of a beam splitter with different splitting ratios
for different polarization components. In Section \ref{sec_BS}
the system is reported which uses a custom-made beam
splitter with similar properties. The phase-covariant
cloner based on state filtration that combines fiber and
bulk optics is described in Section \ref{sec_filt}.
Implementation fully based on fiber optics and spatial mode
encoding is addressed in Section \ref{sec_fibers}. Finally, Section
\ref{sec_concl} summarizes the results and compares
different experimental platforms.

%%%%%%%%%%%%%%%%%%%%%%%%%%%%%%%%%%%%%%%%%%%%%%%%%%%%%%%%%%%%%%%%%%%%%%%%%%%%%%%%%%%%%%%%%%%%%%%%%%%%

\begin{figure}[tbh]
\centerline{\psfig{figure=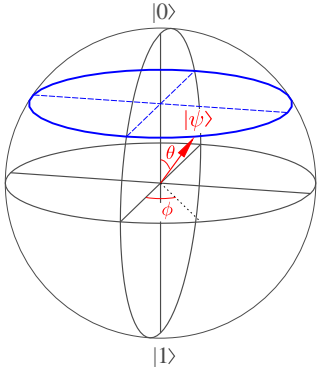,width=0.6\columnwidth}}
\caption{States with the same latitude on the Bloch sphere, e.i. states with fixed 
``intensity'' ratio of basis states $|0\rangle$ and $|1\rangle$. 
$\theta$ and $\phi$ represent the spherical coordinates 
the latitude and longitude, respectively.
\label{Bloch}}
\end{figure}

\section{Theory}\label{Sec_Theory}

In this paper we are interested in optimal copying of single-qubit states $|\psi\rangle$,
which form a two-dimensional Hilbert space spanned by computational basis states 
$|0\rangle$ and $|1\rangle$ and can be parametrized by two angles $\theta$ and $\phi$,
\begin{equation}
|\psi \rangle = \cos{\frac{\theta}{2}} |0 \rangle + e^{i \phi}
\sin{\frac{\theta}{2}} |1 \rangle.
 \label{eq:qubit}
\end{equation}
These states can be visualized as points on the surface of the Bloch sphere, see
Fig.~\ref{Bloch}. Note that $\theta$ and $\phi$ represent the spherical coordinates 
of these points, i.e. the latitude and longitude, respectively.
In certain applications such as eavesdropping on quantum key distribution protocols, one
only needs to copy states that have a fixed latitude $\theta$ on the Bloch sphere. 
In this case the best strategy is to employ an appropriate \emph{phase-covariant} cloner 
that optimally exploits this a-priori information and provides higher
fidelities of the clones than the universal cloner.

The optimal symmetric phase-covariant cloning transformation for the states 
on the northern hemisphere of the Bloch sphere reads
\cite{Niu99,Bruss00,Fiurasek03,Rezakhani05},
\begin{equation}
\begin{array}{ll}
 |0 \rangle \rightarrow & |00 \rangle , \\
 |1 \rangle \rightarrow & \frac{1}{\sqrt{2}} (|10 \rangle + |01 \rangle ).
\end{array}
\label{PC_transformace}
\end{equation}
This transformation produces two approximate clones with identical fidelities.
For the  states from the southern hemisphere the optimal phase-covariant cloning operation
can be obtained simply interchanging the states $|0\rangle$ and $|1\rangle$ in Eq. 
(\ref{PC_transformace}). For the states with the same latitude on
the sphere the cloning fidelities are constant. The minimum value
of fidelity is obtained for the states from the equator of the
Bloch sphere and the fidelities increase up to unity as the signal state
approaches the pole \cite{Fiurasek03}.

In this article we focus on the cloning of the equatorial states
\begin{equation}
 |\psi\rangle= \frac{1}{\sqrt{2}}(|0 \rangle+e^{i\phi}|1 \rangle).
  \label{eq:sym}
\end{equation}
For this class of states the cloning fidelity reaches the value $F_{\rm
ph.cov.} = \frac{1}{2} \left( 1 + \frac{1}{\sqrt{2}} \right)
\approx 85.4\%$. In comparison the fidelity of the optimal symmetric
universal cloning is only $F_{\rm univ} = {5 \over 6} \approx 83.3\%$ and
the semi-classical limit is $F_{\rm sc} = 75\%$. By the semi-classical 
limit we mean the optimal estimation from a single copy of an unknown 
state from the equator of the Bloch sphere \cite{Derka98} followed 
by preparation of two copies according to the measurement result.

\begin{figure}[htb]
\centerline{\psfig{figure=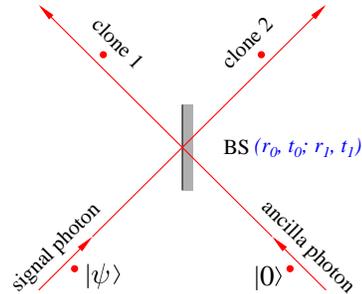,width=0.6\columnwidth}}
\caption{Phase-covariant cloner based on two-photon interference on a 
special beam splitter with splitting ratio 21:79 for the basis state $|0\rangle$ and 
opposite splitting ratio 79:21 for the other basis state $|1\rangle$.
\label{specBS}}
\end{figure}

In most of the experimental schemes presented below we utilize encoding of qubits 
into polarization degree of freedom of single photons. The basis states 
$|0\rangle$ and $|1\rangle$ then correspond to two orthogonal polarization states 
of a single photon, such as vertical, $|V\rangle$, and horizontal, $|H\rangle$, linear 
polarizations. The transformation (\ref{PC_transformace}) can be directly, albeit 
probabilistically, realized by letting the cloned photon interfere with a second
auxiliary photon on a special beam splitter with splitting ratios different 
for state $|0\rangle$, and for state $|1 \rangle$ (or $|V\rangle$ and $|H\rangle$), 
see Fig.~\ref{specBS}. The cloning succeeds when a single photon appears at each 
output port of the beam splitter. The resulting conditional transformation can be 
expressed as
\begin{equation}
\begin{array}{ll}
|0 \rangle _{\rm sig} |0 \rangle _{\rm anc} \rightarrow & (r_0^2 - t_0^2) |00 \rangle , \\
|1 \rangle _{\rm sig} |0 \rangle _{\rm anc} \rightarrow & r_0 r_1 |10 \rangle - t_0 t_1 |01 \rangle , \\
\end{array}
\label{realBS}
\end{equation}
where $r_0, r_1; \, t_0, t_1$ are the amplitude reflectances and transmittances 
for state $|0 \rangle$ and $|1 \rangle$, respectively. We have to find conditions 
under which the action of the beam splitter (\ref{realBS}) corresponds
to the demanded transformation (\ref{PC_transformace}). We immediately find that
the following relations must be fulfilled: 
$r_0r_1= -t_0t_1$ and $(r_0^2 - t_0^2) = \sqrt{2} r_0 r_1$.
%
%%% this yields            $(r_0^2 - t_0^2) = \sqrt{2} r_0 t_0$.
% 
Assuming ideal lossless beam splitter satisfying
$|r|^2 + |t|^2 = 1$ we obtain the value of the intensity reflectances 
$R_0 \equiv r_0^2 = \frac{1}{2}
\left( 1+ \frac{1}{\sqrt{3}} \right)\approx 78.9\%$ and $R_1=1-R_0\approx 21.1\%$.
The multiplicative factors in Eq.~(\ref{realBS}) are related to
the probability of success of the scheme 
$P_{\rm succ1} = (r_0^2 - t_0^2)^2 = (2R_0 -1)^2$, 
which in the ideal case equals to $P_{\rm succ1} = \frac{1}{3}\approx 33.3\%$ \cite{Fiurasek03}.

\begin{figure}[htb]
\centerline{\psfig{figure=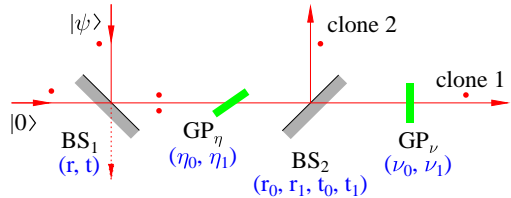,width=0.8\columnwidth}}
\caption{Phase-covariant cloner based on the Hong-Ou-Mandel interference \cite{Hong87} 
on a balanced beam splitter followed by state filtering. \label{BSkloner}}
\end{figure}

%%%%%%%%%%%%%%%%%%%%%%%%%%%%%%%%%%%%%%%%%%%%%%%%%%%%%%%%%%%%%%%%%%%%%%%%%%%%%%%%%%%%%%%%%%%%%%%%%%%%

Another possibility how to obtain the transformation
(\ref{PC_transformace}) for polarization states of photons
is to start from the  implementation of the universal
cloner based on photon bunching on a balanced beam 
splitter \cite{Ricci04,Irvine04,Khan04} and modify it 
by the state filtering, see Fig.~\ref{BSkloner}.
The first beam splitter BS$_1$ implements a universal
cloning. This operation succeeds if both photons leave
BS$_1$ through the same output port. To achieve the optimal
phase-covariant copying operation the state filtering is applied to
the pair of qubits by means of a tilted glass plate GP$_{\eta}$ that 
introduces polarization dependent losses. Finally, the two
photons are separated at the second beam splitter BS$_2$.
The full operation is successful if one photon appears at 
output 1 and simultaneously the other photon at output 2. 
Slight polarization dependence of the transmittance of BS$_2$ 
is compensated by another tilted glass plate GP$_{\nu}$. 
The resulting conditional transformation reads
\begin{equation}
\begin{array}{ll}
|0 \rangle _{\rm sig} |0 \rangle _{\rm anc} \rightarrow
  &  2 r t \eta_0^2   t_0 \nu_0 r_0 |00 \rangle, \\
|1 \rangle _{\rm sig} |0 \rangle _{\rm anc} \rightarrow
  & r t \eta_0 \eta_1 (t_1 \nu_1 r_0 |10 \rangle + t_0 \nu_0 r_1 |01 \rangle ),\\
\end{array}
\label{BS_12}
\end{equation}
where $r,t$ are the coefficients of amplitude reflectances and
transmittances of BS$_1$; $r_0, r_1, t_0, t_1$ are the
coefficients corresponding to BS$_2$ which, in reality,
slightly differ for state $|0 \rangle$ and $|1 \rangle$; 
$\eta_0, \eta_1$ and $\nu_0, \nu_1$ are amplitude transmittances of
the tilted glass plates GP$_{\eta}$ and GP$_{\nu}$, respectively. Again we have
to find conditions under which the transformations (\ref{BS_12})
become equivalent to (\ref{PC_transformace}). In this case, the free parameters
are the transmittances of the tilted glass plates. The conditions for the
symmetric operation read $\nu_0/\nu_1 = t_1r_0/t_0r_1$
and $\eta_1/\eta_0 = \sqrt{2}r_0/r_1$ which determines the tilts of the glass plates. 
The probability of success $P_{\rm succ2} = (2 r t \eta_0^2 t_0 \nu_0 r_0)^2$. Here the 
maximum probability of success reached in ideal conditions is only 
$P_{\rm succ2} = \frac{1}{16} = 6.25\%$.
%
%%%%%%%%%%%%%%%%%%%%%%%%%%%%%%%%%%%%%%%%%%%%%%%%%%%%%%%%%%%%%%%%%%%%%%%%%%%%%%%%%%%%%%%%%%%%%%%%%%%%

\section{Free space realization with the Mach-Zehnder interferometer}\label{sec_MZ}

\begin{figure}[htb]
\centerline{\psfig{figure=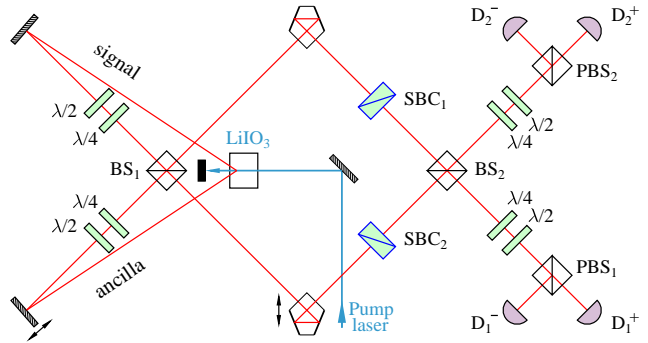,width=1\linewidth}}
\caption{Scheme of the cloning setup based on the Mach-Zehnder interferometer. 
%Experimental setup of a cloner with a Mach-Zehnder interferometer as a beam splitter with
%variable splitting ratio. 
    BS - non-polarizing cube beam splitter, SBC - Soleil-Babinet compensator, 
    PBS - polarizing cube beam splitter, $\lambda/2, \; \lambda/4$ - wave plates,
    D - detector.
\label{setupMZ}}
\end{figure}

In bulk optics it is straightforward to use polarization
encoding of single photon qubits. The states $|0 \rangle$
and $|1 \rangle$ are represented by vertical, $V$, and
horizontal, $H$, linear polarizations, respectively. One
possible way how to simulate a beam splitter with any
required splitting ratio is to use an interferometer and
stabilize it at a certain point within an interference
fringe. The beam splitter with splitting ratios different for
vertical and horizontal polarization can be implemented
by Mach-Zehnder (MZ) interferometer with different phase
shifts for vertical and horizontal polarization components.
Figure \ref{setupMZ} shows the experimental setup utilizing the MZ
interferometer with a Soleil-Babinet compensator in each
arm. With these compensators we can tune the interference
fringes separately for $V$ and $H$ polarizations. In this
setup one must try to carefully compensate
polarization-dependent phase shifts of all used optical
components.

Let us assume the case of perfect balanced MZ interferometer with two 
ideal 50:50 beam splitters without additional phase shifts. The formulas
for the reflectances and transmittances for both polarizations reduce to a simple 
expression
\begin{equation}
R_j = \sin^2{\vartheta_j}, \;\; T_j = \cos^2{\vartheta_j}, \;\;\; j= V,H,
\end{equation}
where $\vartheta_{V}, \vartheta_{H}$ are the phase differences between the MZ interferometer 
arms for the two respective polarizations.

Figure \ref{setupMZ} depicts the whole corresponding experimental cloning setup. The nonlinear 
crystal of LiIO$_3$ is pumped by a cw Kr$^+$ laser at 413 nm. In the crystal, pairs 
of photons are produced in the process of type I spontaneous parametric down-conversion. 
These photons are tightly correlated in time and horizontally polarized. 
Arbitrary polarization state $|\psi\rangle$ of the signal photon 
can be prepared by means of half- and quarter-wave plates ($\lambda/2, \lambda/4$). 
In a similar way, ancilla photon is set to fixed vertically linearly polarized state.
Both photons enter the MZ interferometer, which emulates beam splitter whose 
splitting ratio is independently tunable for horizontal and vertical polarizations.
The MZ interferometer is not perfectly stable. The phase drifts randomly due to 
temperature changes and air flux. Therefore the whole interferometer is enclosed 
in a shielding box and besides that, the interferometer has to be actively stabilized 
during the course of the measurement.

The cloning procedure is successful if there is one photon in each output port. 
The setting of the wave plates at the output ports of the MZ interferometer is 
inverse with respect to the signal qubit preparation, so a photon with the same
polarization as the original signal photon is transmitted through the polarizing 
beam splitter (PBS) to the detector $D^+$, whereas the photon with orthogonal 
polarization is reflected to the detector $D^-$. Each clone is thus measured in 
the basis formed by the input state $|\psi\rangle$ and its orthogonal counterpart. 
In the experiment we measure four coincidence rates $C^{ab}$ of simultaneous 
clicks of detectors $D_1^{a}$ and $D_{2}^b$, where $a,b =+,-$. 
For instance,  $C^{++}$ denotes the number of simultaneous clicks of detectors 
$D_1^+$ and $D_2^+$, which indicates detection of two perfect clones.
The fidelity of the $j$th clone is then calculated as the fraction of the events 
when detector $D_j^{+}$ fired and the total number of all detection events,
\begin{eqnarray}
  F_1 &=& \frac{C^{++} + C^{+-}}{C_{\mathrm{sum}}} ,\nonumber \\
  F_2 &=& \frac{C^{++} + C^{-+}}{C_{\mathrm{sum}}} ,
\end{eqnarray}
where $C_{\mathrm{sum}} = {C^{++} + C^{+-} + C^{-+} + C^{--}}$.

In order to correctly measure $C^{ab}$ we must ensure that the detection 
efficiencies of all four detectors are identical, otherwise we could obtain 
biased results. The relative efficiencies can be balanced by several ways. 
First, we can measure exact efficiency of each detector and then calculate the
fidelity multiplying the measured coincidences accordingly. Second, we can add 
additional losses in front of each detector (reduce the iris diameter) 
to balance the efficiencies. Third, we can change the measurement basis and measure 
all four coincidences in a sequence using only two detectors, therefore it is not 
necessary to compensate for any differences. We checked that all three methods 
provide the same results.

\begin{figure}[hbt]
\centerline{\psfig{figure=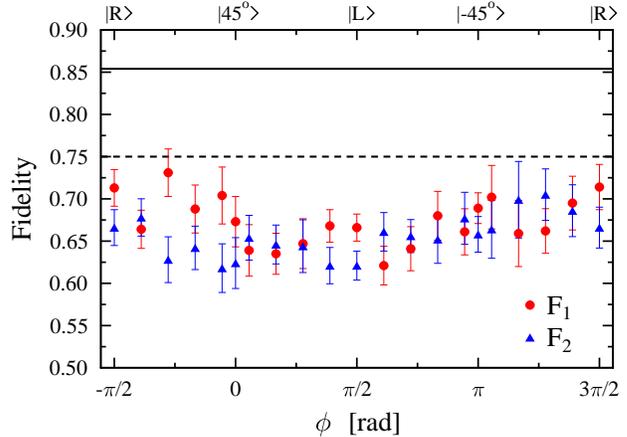,width=1\linewidth}}
\caption{Fidelities of clones measured with the setup based on the MZ interferometer. 
         Full line denotes the theoretical value of fidelity of the phase-covariant cloner, dashed  
         line shows the semi-classical limit. 
\label{fidMZ}}
\end{figure}
Figure \ref{fidMZ} shows the measured fidelities of clones of the states from 
the equator of the Bloch sphere. Despite of relatively precise adjustment, the 
fidelities are below the semi-classical limit. This is caused mainly by the 
imperfect overlap of spatial modes on the beam splitters, and by the intrinsic 
phase shifts of the cube non-polarizing BSs, which are different for reflected
and transmitted photons and cannot be fully compensated in our experiment.
Partial compensation is possible for some subset of input states, but complete 
compensation is experimentally unfeasible. Finally due to the interferometric 
setup, the phases may slightly drift in the course of measurement.

The probability of success is determined as the ratio of the sum of all coincidence 
events at the output of the device and the number of photon pairs entering the MZ 
interferometer.
The average probability of success, $P_{\rm succ} = 33.3 \pm 0.2 \%$, corresponds 
well to the theoretical value.

The main advantage of this setup is the possibility to set any splitting 
ratio which is essential for asymmetric cloning with tunable asymmetry 
\cite{Bartuskova06}. The main disadvantages include uncontrollable 
phase shifts that could not be compensated and non-perfect overlap 
of the spatial modes on the beam splitters. Consequently, the cloning fidelities
were rather low and could not be improved even if we performed spatial
mode filtering by single-mode fibers between the nonlinear crystal and the interferometer.

%%%%%%%%%%%%%%%%%%%%%%%%%%%%%%%%%%%%%%%%%%%%%%%%%%%%%%%%%%%%%%%%%%%%%%%%%%%%%%%%%%%%%%%%%%%%%%%%%%%%

\section{Free space realization with special beam
splitter}\label{sec_BS}

\begin{figure}[htb]
\centerline{\psfig{figure=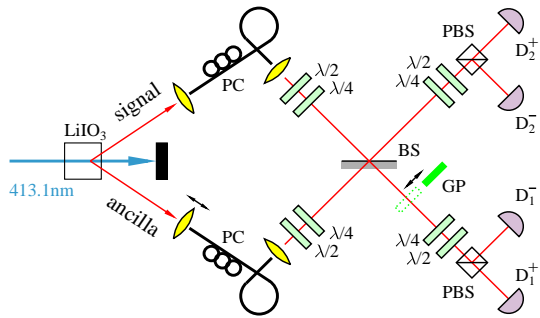,width=.85\linewidth}}
\caption{Scheme of the cloning setup based on the specially fabricated beam splitter.
         BS - special unbalanced plate beam splitter, PBS - polarizing cube beam splitter, 
         PC - polarization controller, GP - compensation glass plate, 
         $\lambda/2, \; \lambda/4$ - wave plates, D - detector.
\label{setup80_20}}
\end{figure}

\begin{figure}[bth]
\centerline{\psfig{figure=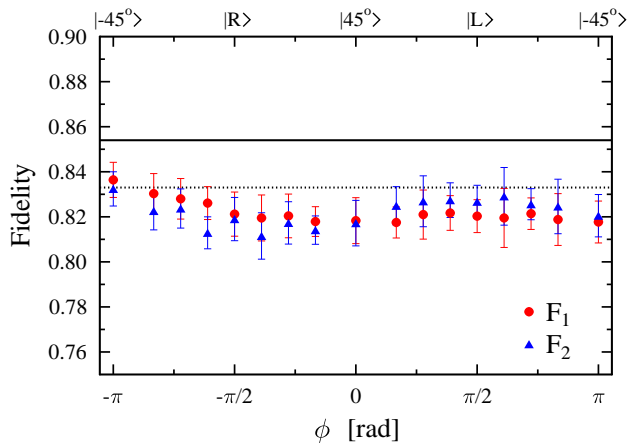,width=1\linewidth}}
\caption{Fidelities of clones measured with the setup based on the special beam splitter together
         with a compensation glass plate. Full line denotes the theoretical value of fidelity of 
         the phase-covariant cloner, dotted line shows the limit of the universal cloner.
\label{fid80_20}}
\end{figure}

If the tunability of the splitting ratio is not desired then an alternative setup with a
fixed beam splitter instead of the MZ interferometer is a good choice. 
We utilized a special beam splitter (manufactured by Ekspla) for this purpose 
and built a new setup as displayed in Fig.~\ref{setup80_20}. Down-converted 
photon pairs from the source are coupled into the single-mode fibers, released back
into free space and then they enter the cloner. The fibers ensure precise spatial mode filtering
that enhances the overlap of beams on the bulk beam splitter. 
Because the actual splitting ratios of the BS are not
exactly 21:79 for vertical polarization and 79:21 for horizontal polarization 
as required by the theory (given in Sec.~\ref{Sec_Theory}) we compensated 
them by a tilted glass plate GP. For comparison, we made two series of measurements. 
The first series was made without any compensation and the
measured fidelities of two clones differed by several percent,
$F_1 = 84.1 \pm 0.2 \%$ and $F_2 = 80.4 \pm 0.2 \%$. The probability of success was 
$P_{\rm succ} = 31.23 \pm 0.08 \%$.
The second series of measurements was made introducing polarization dependent 
losses in one output by the tilted GP. In this case we achieved more symmetric 
operation, when the measured fidelities became equal within the
measurement error $F_1 = F_2 = 82.2 \pm 0.2 \%$ as 
displayed in Fig.~\ref{fid80_20}. However, these
additional losses slightly decreased the probability of success,
$P_{\rm succ} = 28.8 \pm 0.1 \%$.

The main advantage of this experimental scheme is its simplicity. It essentially 
generalizes the Hong-Ou-Mandel interferometer \cite{Hong87} using special 
unbalanced beam splitter. The setup is very stable and compact, losses are minimal
and it is not required to actively stabilize any second order interference.
The spatial mode filtering by optical fibers increases the overlap of the spatial 
modes of signal and ancilla photons. Nevertheless, the visibility of HOM-type
interference is still not perfect. Using balanced 50:50 beam splitter we typically 
reach values $\approx 92\%$. Consequently, the averaged fidelities of the clones 
do not exceed the universal cloning limit. The main disadvantage of this setup 
is that it is possible to tune the asymmetry of the cloner only by applying additional
losses which decreases the probability of success. More detailed description of 
this particular setup can be found in Ref. \cite{Cernoch06}.

%%%%%%%%%%%%%%%%%%%%%%%%%%%%%%%%%%%%%%%%%%%%%%%%%%%%%%%%%%%%%%%%%%%%%%%%%%%%%%%%%%%%%%%%%%%%%%%%%%%%

\section{Hybrid setup}\label{sec_filt}

\begin{figure}[htb]
\centerline{\psfig{figure=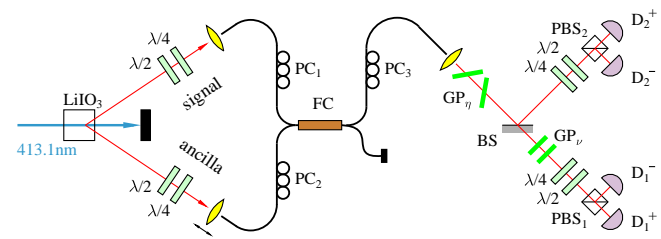,width=1\linewidth}}
\caption{Scheme of the hybrid cloning setup. BS - non-polarizing beam splitter,
         PBS - polarizing beam splitter, PC - polarization controller, FC - fiber coupler,  
         $\lambda/2, \; \lambda/4$ - wave plates, GP$_{\eta}$, GP$_{\nu}$ - glass plates, 
         D - detector.
\label{setupHyb}}
\end{figure}

In order to further increase the cloning fidelities and exceed the fidelity of optimal 
universal cloner, $F_{\rm univ} = {5 \over 6} \approx 83.3\%$, we have experimentally 
implemented the cloning scheme based on photon bunching and state filtering 
depicted in Fig.~\ref{BSkloner}. The resulting hybrid setup  combines advantages 
of fiber and free-space approaches, and is schematically sketched in Fig.~\ref{setupHyb}. 
The signal and ancilla photons are coupled into single-mode fibers and interfere in 
a fiber coupler (FC). Coupling the photon pairs into fibers selects only very well
defined spatial modes of the down-converted field. The spatial wavepackets of the photons
perfectly overlap and the actual splitting ratio of the fiber coupler is 49:51. 
These conditions guarantee very high visibility of the HOM interference,
typically $\approx 98\%$. The free-space part allows easy handling of the
information encoded in the polarization of the photons.

The measurement starts with adjusting the
HOM dip. Both fiber outputs from the FC are connected
directly to the detectors and the path difference is
adjusted in order to maximize the fourth order interference,
i.e., to minimize the coincidence counts. The polarization
changes caused by the fibers are compensated using the
polarization controllers (PC$_1$ and PC$_2$).

Only single output of the FC is used in the cloning
operation. In fact, FC implements optimal universal cloner that is 
converted to the phase-covariant one by means of state
filtering provided by the glass plate GP$_{\eta}$ that
introduces polarization dependent losses. After this filtering step, 
the non-polarizing beam splitter BS splits the two photons into two 
different paths with probability ${1 \over 2}$. The polarization dependence of
the splitting ratio of this BS is compensated by polarization
dependent losses introduced by the glass plate GP$_{\nu}$.
Then the polarization analysis is performed in a standard
way as in the previous setups. The setting of polarization
states of signal and ancilla photons is here more complicated than 
in the previous cases. First we tilt the PC$_3$ to adjust the linear 
vertical polarization of ancilla photon at the input of BS, while all 
the wave plates are rotated to zero position. After that the ancilla 
photon is effectively linearly vertically polarized. 
The particular signal states are prepared in the similar way: 
The ancilla arm is blocked and the measurement polarization bases
in the detection blocks are set to select a required polarization 
state. Then the input polarization is tuned so that to reach 
the situation when all the signal photons are detected on the 
detectors $D^+$ only.

\begin{figure}[b]
\centerline{\psfig{figure=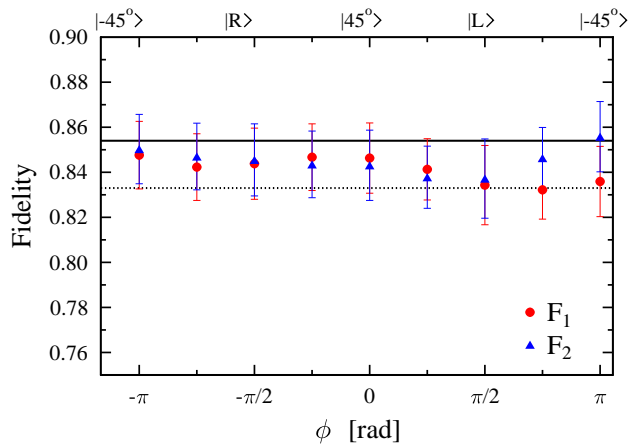,width=1\linewidth}}
\caption{Fidelities of clones measured with the hybrid setup. Full line denotes the theoretical 
         value of fidelity of the phase-covariant cloner, dotted line shows the limit of the 
         universal cloner.
\label{fidHyb}}
\end{figure}

As noted above, without GP$_{\eta}$ (without the polarization filtration 
in front of BS) the setup operates as a well known universal cloning machine 
\cite{Ricci04,Irvine04,Khan04}. Note that the
overlap of the vertically polarized ancilla photon with any
equatorial state of the signal photon is always ${1 \over 2}$.
For comparison we measured also the fidelities of this universal cloning
operation for the same set of input states (not shown here). 
The mean values of fidelities with this universal cloner were 
$F_1 = 82.5 \pm 0.9\%$ and $F_2 = 82.5 \pm 0.6\%$, which are very close to the 
theoretically expected $F_{\rm univ} \approx 83.3\%$. The fidelities measured 
with the phase-covariant cloner are showed in Fig.~\ref{fidHyb}. 
With this device we finally exceeded the universal cloner limit, and achieved mean 
fidelities $F_1 = 84.1 \pm 0.6\%$ and $F_2 = 84.5 \pm 0.6\%$.

To summarize, the main advantage of the hybrid setup is that it is easy
to achieve high visibilities and exceed the universal cloning limit. The 
disadvantage is lower probability of success of this cloning scheme, theoretical 
maximum is ${1 \over 16}=6.25\%$. Due to the losses introduced mainly by the 
compensating glass plates we obtained $P_{\rm succ} = (4.2 \pm 0.1) \%$.
It is also more difficult to properly set the signal and ancilla photon polarization.

%%%%%%%%%%%%%%%%%%%%%%%%%%%%%%%%%%%%%%%%%%%%%%%%%%%%%%%%%%%%%%%%%%%%%%%%%%%%%%%%%%%%%%%%%%%%%%%%%%%%

\section{Fiber setup}\label{sec_fibers}

\begin{figure}[htb]
\centerline{\psfig{figure=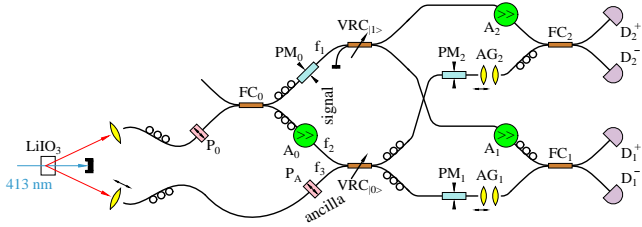,width=1\linewidth}}
\caption{Scheme of the all-fiber cloning setup: 
         P - polarizer, FC - fiber coupler, A - attenuator, PM - phase modulator,
         VRC - variable ratio coupler, AG - air gap, D - detector.
\label{setupFib}}
\end{figure}

The last setup is completely composed of fibers and fiber optics components. 
The polarization of photon propagating through the fiber undergoes transformations
that are hard to trace. Moreover, some fiber components in this setup transmit 
only one polarization. This makes the encoding of qubits into polarization states 
in all-fiber scheme inconvenient. However, our setup utilizes spatial-mode
encoding of qubit states (see Fig.~\ref{setupFib}). Each qubit is represented 
by a single photon which can propagate through two different fibers. The presence 
of the photon in the first (second) fiber corresponds to the basis state 
$|0 \rangle$ ($|1 \rangle$). The intensity ratio and phase difference between 
these two modes determine the state of the qubit, Eq. (\ref{eq:qubit}).

Signal and ancilla photons are created by the same source of down-converted 
photon pairs as mentioned above. The signal photon is split by a fiber coupler 
FC$_0$ into fibers $f_1$, corresponding to the basis state $|1 \rangle$, and $f_2$, 
corresponding to state $|0 \rangle$. With this setup we have experimentally realized 
the cloning of equatorial qubit states, Eq. (\ref{eq:sym}). Therefore unequal losses 
in optical fibers $f_1$ and $f_2$ are balanced using the attenuator A$_0$. Various 
states from the equator are prepared by applying appropriate voltage to the phase 
modulator PM$_0$, which sets the relative phase $\phi$. The ancilla photon is always 
in the fixed state $|0 \rangle$ corresponding to the single photon propagating 
exclusively through fiber $f_3$.

The cloning procedure is accomplished by two variable-ratio couplers VRC$_{|0 \rangle}$ 
and VRC$_{|1 \rangle}$ where the first one forms the core of the HOM interferometer. Before 
starting the final measurement, the splitting ratio of VRC$_{|0 \rangle}$ is set to 
50:50 and the HOM dip is adjusted similarly as in the case of the hybrid setup. The 
visibility of the fourth order interference is typically $\approx 98\%$. Then the splitting 
ratio of VRC$_{|0 \rangle}$ is changed to the desired value 79:21, whereas the reverse 
splitting ratio 21:79 is set on VRC$_{|1 \rangle}$.

Both MZ interferometers are adjusted using only the signal
beam from the nonlinear crystal. First the intensities
between arms of each MZ interferometer are balanced with the
help of attenuators in the detection part of the setup (the
part behind VRCs). Visibilities are maximized by precisely balancing 
the optical lengths in both arms and aligning polarizations in each 
interferometer. We typically reached visibilities of the second order
interference about $97\%$. After this we  set damping factors in the 
detection parts such as to ensure projections onto the states on the 
equator of the Bloch sphere.

Fluctuations of temperature and temperature gradients cause a random phase drift between arms of
each MZ interferometer. Therefore the experimental setup is thermally isolated in a polystyrene box
and additionally an active stabilization is periodically performed between three-second measurement
steps. Only signal beam from the crystal is used for the stabilization, the other one is blocked.
In each stabilization cycle values of both phase drifts are simultaneously estimated and they are
compensated by means of phase modulators in the detection part of the setup.

\begin{figure}[htb]
\centerline{\psfig{figure=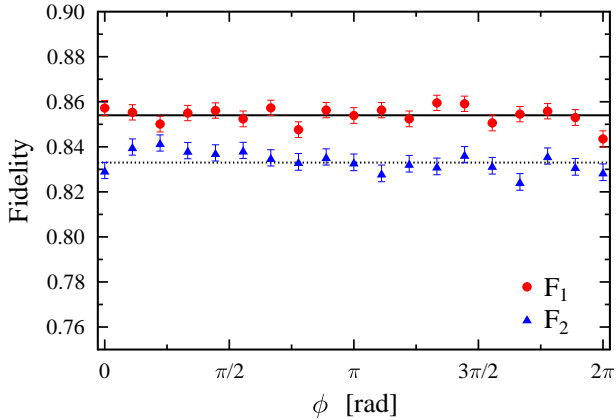,width=1\linewidth}}
\caption{Fidelities of clones measured with the all-fiber setup. Full line denotes the theoretical 
         value of fidelity of the phase-covariant cloner, dotted line shows the limit of the 
         universal cloner.
\label{fidFib}}
\end{figure}

The cloning operation is successful only if there is one photon detected 
by each pair of detectors $(D^{+}, D^{-})$. Fidelities of the clones are 
measured using two detection blocks consisting of the phase modulators,
the attenuators and 50:50 fiber couplers. The projection onto the particular 
signal state is realized by the setting of proper phase shifts by phase modulators 
PM$_1$ and PM$_2$. Unequal detector efficiencies are compensated by proper 
rescaling of the measured coincidence rates according to the relative detector 
efficiencies. We also checked, that the same fidelities are obtained from 
the coincidences measured only by one pair of detectors. Acquired fidelities 
are shown in Fig.~\ref{fidFib}, their values averaged over all measured
states are $F_1 = 85.4 \pm 0.4 \%$ and $F_2 = 83.4 \pm 0.4\%$. They differ by 
2.0\%, which is probably the consequence of non-ideal splitting ratios, 
differing from 50:50, of the couplers FC$_1$ and FC$_2$ in the detection blocks. 
Also the fidelity $F_2$ is more sensitive to proper adjustment of the HOM dip 
due to the unbalanced splitting ratio of VRC$_{|0 \rangle}$.

Our experiment demonstrates that the fiber optics enables us to reach 
high interference visibilities and achieve fidelities exceeding the 
universal cloning limit. This setup is compatible with fiber-based 
communications and can be also used as an asymmetric phase-covariant cloner 
simply by changing the splitting ratios of VRCs, see Ref. \cite{Bartuskova06}. 
However, some fiber components cause significant power losses. Although the
probability of success of the cloning operation itself is relatively high, 
$P_{\rm succ} = (33.5 \pm 0.3) \%$, the actual cloning rate is very low, 
in our case about $60$ per second, due to losses in the state-preparation 
and detection part of the setup. 

%%%%%%%%%%%%%%%%%%%%%%%%%%%%%%%%%%%%%%%%%%%%%%%%%%%%%%%%%%%%%%%%%%%%%%%%%%%%%%%%%%%%%%%%%%%%%%%%%%%%

\section{Conclusions}\label{sec_concl}

We constructed four experimental setups realizing symmetric phase-covariant cloning 
of single photon qubits. The information is encoded either into polarization, or into 
spatial modes. The cloning operation is implemented probabilistically by interference of 
the signal photon with an ancilla photon. Three setups realized the cloning by 
using a special beam splitter with different transmittance for the two basis states
$|0\rangle$ and $|1\rangle$. In contrast, the hybrid setup realized the cloning by an
alternative approach employing two standard 50:50 beam splitters followed by state 
filtration introduced by polarization dependent losses.

\begin{table}[h]
\centerline{
\begin{tabular}{|l|c|c|c|} \hline
cloner type  & $F_1$   & $F_2$   & $C_{\mathrm{sum}}$ \\ \hline
Mach-Zehnder & 67 \%   & 66 \%   & 1700 /s \\ \hline
special BS   & 82.2 \% & 82.2 \% & 3780 /s \\ \hline
Hybrid       & 84.1 \% & 84.5 \% & 680 /s \\ \hline
Fiber        & 85.4 \% & 83.4 \% & 60 /s \\ \hline
\end{tabular}}
\caption{Compared values of fidelities $F_1$, $F_2$ and total clone rates 
         $C_{\mathrm{sum}}$ for all four devices.\label{nase:tabulka}}
\end{table}

Table \ref{nase:tabulka} summarizes our measurement results. 
The first setup represents 
purely free space realization based on the Mach-Zehnder interferometer. This setup is very
flexible, any required transmittance can be adjusted by tuning the Soleil-Babinet 
compensators. However, imperfections of individual components sum up so that
the average cloning fidelity is bellow the semi-classical limit. 
The simplification of this cloning setup led to the second setup based on the specially designed
beam splitter. The average cloning fidelity is close to the limit of the universal cloner. 
The asymmetry of the clones is not easily tunable. The main advantage is the high 
coincidence rate.
We exceeded the limit of the universal cloner for cloning of polarization states with 
the third setup. It takes advantages of the perfect overlap of spatial modes in the 
fiber based beam splitter (the fiber coupler) and it also uses simple encoding into 
polarization modes. The main disadvantage of this strategy is smaller theoretically 
attainable probability of success.  
The last setup is composed of fiber components exclusively. The fidelities measured 
with this setup also exceeded the universal cloning limit. In this case, the qubits 
were encoded into spatial modes. This type of encoding permits easy tunability 
of the asymmetry, but the setup composed of two Mach-Zehnder interferometers is rather 
sensitive to fluctuations and requires active stabilization. 

\section{ACKNOWLEDGEMENTS}

We thank Eva Kachl\'{i}kov\'{a} for her help in the preparatory stage of the experiment. 
This research was supported by the projects LC06007, 1M06002 and MSM6198959213 
of the Ministry of Education of the Czech Republic and by the EU project SECOQC 
(IST-2002-506813).

\end{document}